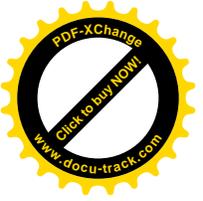
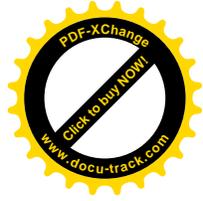

# Bloch vector analysis in non linear, finite, dissipative systems: an experimental study

G. D'Aguanno[1,2*], M.C. Larciprete[3], N. Mattiucci[1,2], A. Belardini[3], M.J. Bloemer[2],

E. Fazio[3], O. Buganov[4], M. Centini[3], and C. Sibilia[3].

[1]AEgis Tech., Microsytems/Nanotechnology, 631 Discovery Drive Huntsville AL

35806, USA

[2] C. M. Bowden Facility, Bldg 7804, RDECOM , Redstone Arsenal AL35898, USA

[3] CNISM and Dipartimento di Energetica, Sapienza Università di Roma

Via A. Scarpa 16, I-00161 Roma, ITALY

[4] Institute of Molecular and Atomic Physics, NASB, Minsk BY-220072, Belarus

**Abstract**

We have investigated and experimentally proved the robustness of the Bloch vector for one-dimensional, nonlinear, finite, dissipative systems. The case studied is the second harmonic generation from metallo-dielectric filters. Nowadays metallic based nanostructures play a fundamental role in nonlinear nano-photonics and nano-plasmonics. Our results clearly suggest that even in these forefront fields the Bloch vector continues to play an essential role.

[*] Corresponding Author: giuseppe.daguanno@us.army.mil Tel+1-256-842-9815 Fax+1-256-842-2507

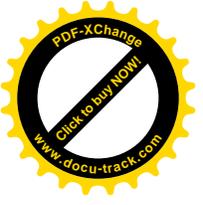
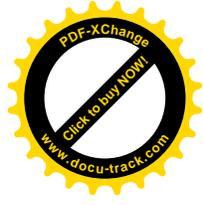

*Introduction*-The Bloch theorem and the Bloch vector are protagonists in many fields of physics ranging from solid state physics [1] to optics and photonics [2]. Anywhere the periodic repetition of elements in one, two or three dimensions (1-D, 2-D, 3-D) -be they atoms, molecules, thin layers of materials or any generic building block- gives rise to allowed and forbidden bands for wave propagation, there the Bloch vector comes to play as the leading role actor. It would be a tantalizing effort just to cite the many physical phenomena whose interpretation is based on the Bloch vector analysis and clearly the space of this Letter would not be enough. Suffice here just to cite perhaps the most known phenomenon: the electron levels in periodic potential and the electric conduction theory [1,2]. In this Letter we focus on simple 1-D periodic systems (layered structures or Bragg gratings [2]) because of their intrinsically simpler fabrication procedures and easier theoretical analysis than multidimensional systems, although they retain many of the characteristics of more complicated structures. A Bragg grating or 1-D photonic crystal (PC) in its simplest realization is basically made of a periodic repetition of two layers of materials with different refractive indices which form the elementary cell (or building block) of the structure. This periodic repetition gives rise to allowed and forbidden bands for light propagation [2] in analogy with the allowed and forbidden bands for electronic propagation in semiconductors [1]. We can distinguish purely dielectric Bragg gratings and metallo-dielectric (MD) structures, both can be fabricated by standard sputtering or thermal evaporation techniques [3]. The Bloch vector for a 1-D, periodic structure comes directly from the Bloch theorem and can be written as [4]:

$$K_\beta(k_x, \omega) = \frac{1}{\Lambda} \cos^{-1}\left[\frac{1}{2}(m_{11} + m_{22})\right] \qquad (1)$$

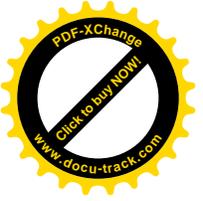
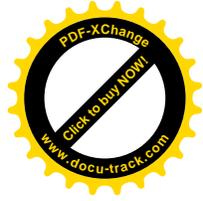

where $cos^{-1}$ is the inverse cosine (arccosine) function, $k_x$ is the transverse wave-vector along the x-axis, i.e. in the direction perpendicular to the periodicity (z-axis), $\Lambda$ is the length of the elementary cell of the structure and $\hat{M} = (m_{1,1}, m_{1,2}; m_{2,1}, m_{2,2})$ is the transfer matrix (or scattering matrix) of the elementary cell. Eq.(1) comes from the Bloch theorem which is applicable for a *strictly* periodic potential. One may ask what happen when the structure under investigation is made just of few periods so that it is intrinsically finite and periodic boundary conditions cannot be applied (finite structure). And moreover, what happen when the materials involved are dissipative. These deceptively simple questions have been actually the subject of an intense theoretical investigation over the last two decades [5-6]. One simple way to approach the problem is the following: we note that a Bloch vector *defined* as in Eq.(1) can be always calculated, regardless of the strictly applicability or not of the Bloch theorem, because it depends on the trace of the scattering matrix of the elementary cell. In other words, Eq.(1) can be calculated no matter the number of the elementary cells that actually compose the structure and no matter if the structure is dissipative or not. Of course, the question that arises is whether or not this *"generalized Bloch vector"* continues to give useful information, especially when, as in our case, one deals with structures of only few periods and strong dissipation. In order to shed some light on the question we have decided to study a second harmonic generation (SHG) process from three MD structures and exploited the possibility to interpret the results in the frame-work of a Bloch vector analysis.

*Samples preparation, SHG model and experimental results*- The three samples (sample a), sample b) and sample c)) are made of N=5 periods of alternating layers of Ag and Ta$_2$O$_5$. The elementary cells of samples are respectively: a) [Ag(~21nm)/Ta$_2$O$_5$(~122nm)], b) [Ag(~18nm)/Ta$_2$O$_5$(~152nm)], c) [Ag(~18nm)/Ta$_2$O$_5$(~169nm)]. Note that in all three structures the amount of Ag is roughly the same. All depositions were carried out by magnetron sputtering onto 1 mm thick, optically flat (λ/20) glass substrates. After deposition was completed, a linear optical characterization of the samples was carried out. The transmittance spectra were recorded at normal incidence in the visible-NIR range by spectro-photometric technique and the experimental curves were reconstructed using a standard transfer-matrix algorithm [4]. The optical constants for Ag used to fit the data were taken from the book of Palik [7], and the optical constants for Ta$_2$O$_5$ were taken from previously measured data [8] of the reflectance of a single Ta$_2$O$_5$ film deposited on Si with a Filmetrics reflectometer having a lower wavelength range of 600 nm. In particular we take the relative permittivity of the materials as follows: $\varepsilon_{Ag}(800nm) = -27.95 + 1.52i$, $\varepsilon_{Ag}(400nm) = -3.77 + 0.67i$, $\varepsilon_{Ta2O5}(800nm) = 4.6 + 0.027i$ and $\varepsilon_{Ta2O5}(400nm) = 4.84 + 0.088i$. Note that all the materials have an imaginary part of the permittivity and therefore they have some degree of dissipation both at the fundamental frequency (FF) and the SH frequency. In our samples the only source of quadratic nonlinearity is the one associated with the metal layers. We measured the reflected SH signal at 400nm for different polarization state of the fundamental input beam, different intensities and incident angles. The fundamental beam was provided by the output of a femtosecond Ti:Sapphire laser (λ=800 nm, 1 kHz repetition rate, 150 fs pulse width), focused close to the sample with lens of a 150 mm focal length. The sample was placed on a rotational stage which allowed setting of the incidence angle, with a resolution of 0.5 deg. The transverse profile of the fundamental beam was measured to be Gaussian with a spot size *w* of ~600 µm corresponding to a peak power of ~6GW/cm$^2$. Fundamental and generated beam polarization states can be selected by rotating a half-wave plate and a linear analyzer, respectively. A long pass filter was used after the half-wave plate in order to avoid the spurious SH signal produced by the plate's crystals itself, due to the short pulse duration. After being reflected by the sample, the fundamental beam was suppressed, thus ensuring that only the SH beam was directed to the photomultiplier tube, and then analyzed by a 500 MHz digital oscilloscope. The calibration curve of the photomultiplier response was accurately performed with a reference BBO crystal. Experimental measurements performed under different polarization state of the FF show that the largest signal is recorded when the polarization of fundamental beam is set to $\hat{p}$ (TM), while the SH signal is $\hat{p}$-polarized for both $\hat{s}$ (TE) and $\hat{p}$ fundamental beam polarization, as expected [9]. This first set of measurements on all the samples was done by increasing the FF peak power and verified the quadratic dependence of the SH signal on the FF peak power. Before going to describe in more details the experimental results we would like to spend few words on the theoretical model used.

The theoretical model to explain the SHG in the MD structure follows the classical approach outlined by Shen [10]. The quadratic nonlinearity of metals is described through two terms: the Lorentz term and the surface term. The Lorentz term is $2i\omega\gamma(z)\left(\vec{E}_\omega \times \vec{H}_\omega\right)$ where $\gamma(z) = \gamma$ in the metal layers and $\gamma(z) = 0$ in the dielectric layers. As from its name, it accounts for the Lorentz force exerted on the free electrons of the metal. The surface term is $d_S^{(2)} \sum_k \delta(z_k) : \vec{E}\vec{E}$ where $\delta(z_k)$ is the Dirac "delta" function calculated at the *k-th* metal/dielectric interface just inside the metal. It accounts for the second order susceptibility at each metal/dielectric interface due to symmetry breaking. For TM polarization, in a Cartesian, right-handed, reference system (x,y,z) where z is the direction of the stratification of the structure, considering only the (z,z,z) component of the nonlinearities, i.e. TM→TM SH emission, the Helmholtz equation in MKSA units for the SH H-field polarized along the y-axis can be written as:

$$\frac{d^2 H_{y,2\omega}}{dz^2} + \frac{4\omega^2}{c^2}\left(n_{2\omega}^2(z) - n_{in}^2 \sin^2 \vartheta\right) H_{y,2\omega} = \frac{4\omega^2 n_{in}}{c} \sin \vartheta \left[ \varepsilon_0 d_s^{(2)} \sum_k \delta(z_k) E_{z,\omega}^2 + \frac{2i\omega}{c^2} \gamma(z) E_{x,\omega} H_{y,\omega} \right]$$

(2)

where $\vartheta$ is the incident angle of the pump field on the sample, $n_{in}$ is the refractive index of the incident medium (air in our case), $n_{2\omega}(z)$ is the step-varying, complex refractive index at the SH along the direction of the stratification, $\varepsilon_0 \cong 8.85 \times 10^{-12}$ F/m is the vacuum permittivity and, obviously, $c$ is the speed of light in vacuo. $E_{x,\omega}$ is the x-component of the FF electric field while $H_{y,\omega}$ is the FF magnetic field. In our model the two "free parameters" are $\gamma$ and the surface nonlinearity $d_S^{(2)}$. We suppose that FF field remains undepleted allowing therefore its calculation through a standard, linear matrix transfer technique [4]. In the undepleted regime, Eq.(2) can be solved using a Green function approach for multi-layered structures developed in Ref.[11]. Note that in the Lorentz term of Eq.(2) we have included only the component oriented along the z-direction as it results the dominant one according to experiments [12].

The experimental data of the conversion efficiency of the reflected SHG versus incident angle $\vartheta$ for the three samples are reported in Figure 1 where it is also reported the comparison with the theoretical predictions. The polarization direction of both fundamental and SH generated electric fields lay in the plane of incidence (TM→TM).

For all the investigated multilayer structures, the SH signal displays a maximum value at an incidence angle of ~55 deg, instead of ~70 deg which is expected for the single Ag layer [9]. The figure also shows the estimated values of the nonlinearity for the three samples. We note that sample a) and sample b) have approximately the same values of nonlinearity although the maximum conversion efficiency of sample a) is one order of magnitude greater than the maximum conversion efficiency of sample b). We also note that sample c) appears to have much smaller values of the nonlinearities with respect to the first two samples.

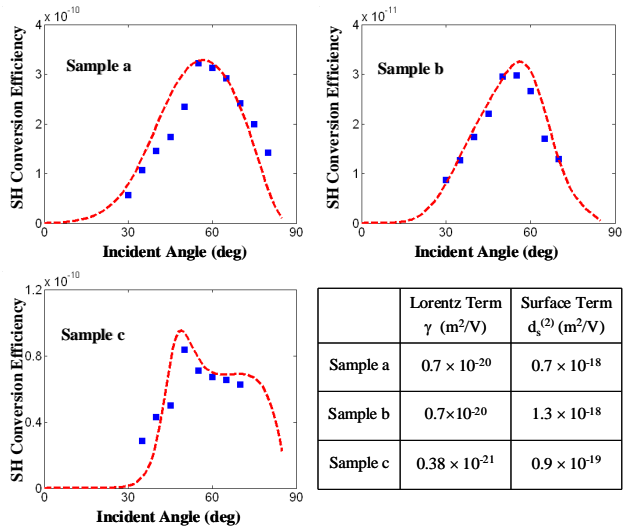

**Figure 1:** (Color online) Reflected SH conversion efficiency vs. incident angle for TM→TM emission, theory (dashed line) and experiment (squares) for each sample. The incident intensity is ~6GW/cm$^2$. Table: Estimated values of the quadratic nonlinearities.

We would like to point out that in our approach the parameters $\gamma$ and $d_S^{(2)}$ must be intended as "phenomenological parameters" in some way similar to those introduced in Ref. [13]. Nevertheless, by considering an effective component of the (z,z,z) nonlinearity $d_\perp^{(2)} \approx (1/2)\left(\gamma + d_s^{(2)}\right)$ as the dominant nonlinearity for TM→TM emission, in agreement with the experimental results of Ref.[12], we found that the three samples have respectively the following values of effective nonlinearity: $d_\perp^{(2)} \approx 3 \times 10^{-19} m^2/V$ (sample a)), $d_\perp^{(2)} \approx 6 \times 10^{-19} m^2/V$ (sample b)), $d_\perp^{(2)} \approx 5 \times 10^{-20} m^2/V$ (sample c)). Those values are, everything considered, in good agreement with the values experimentally measured for silver in Ref.[12] where the data reported for four different samples made of a single Ag layer deposited on a Si substrate range from a minimum of $d_\perp^{(2)} = (\chi_\perp^{(2)}/2) \approx 10^{-20} m^2/V$ to a maximum of $d_\perp^{(2)} = (\chi_\perp^{(2)}/2) \approx 10^{-19} m^2/V$.

*Bloch vector analysis-* We now proceed to the interpretation of the results in the framework of a Bloch vector analysis. At this end, in Fig. 2 we show the Bloch vector as defined in Eq.(1) and the linear transmittance for the three samples. Both quantities are calculated for TM polarization and an incident angle of 55 deg that is approximately the angle where the SH emission shows its maximum for all the samples. First thing we note is that, differently from non dissipative structures, the Bloch vector has an imaginary part even outside the band gaps. We also note that the real part of the Bloch vector (continuous line) closely resembles that one associated with an ideal non dissipative structure. We remark once again that clearly the Bloch theorem does not apply to such structures due to the dissipation and to the limited number of periods, but still a Bloch vector can be defined according to Eq.(1). Now it raises the question whether or not be this generalized Bloch vector sturdy enough to continue to give useful information on the physical phenomenon investigated even in these extreme circumstances.

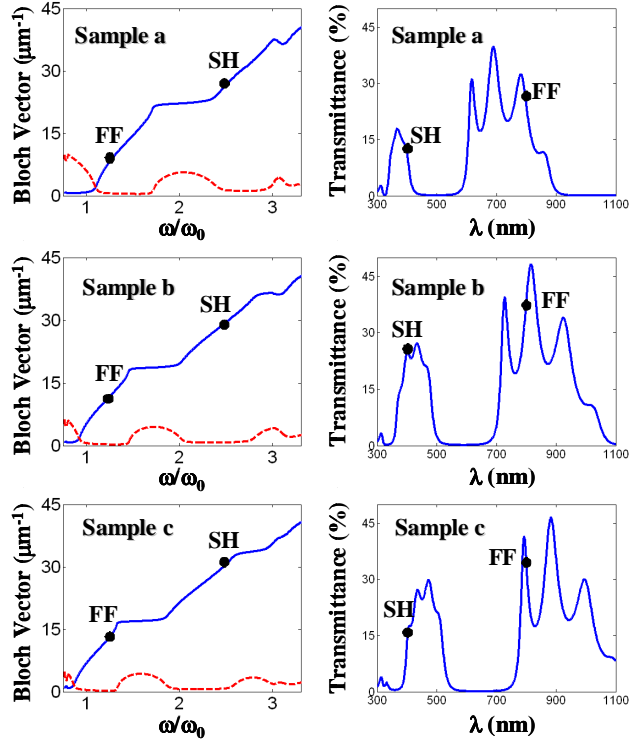

**Figure 2:** (Color online) Left column: Real part (continuous line) and imaginary part (dashed line) of the Bloch vector vs. $\omega/\omega_0$ for the three samples. $\omega_0$ is a reference frequency corresponding to a wavelength of 1µm. Right column: Linear Transmittance vs. wavelength for the three samples. Both quantities are calculated for TM-polarization and an incident angle of 55 deg.

In Table 1 we summarize the values of the real part of the Bloch vector at the FF and SH frequency and the first reciprocal lattice vector $G_1$ for the three samples.

**Table 1**

|  | $ReK_\beta(FF)$ (µm$^{-1}$) | $ReK_\beta(SH)$ (µm$^{-1}$) | $G_1$ (µm$^{-1}$) |
|---|---|---|---|
| Sample a | 8.1 | 26.7 | 43.6 |
| Sample b | 11.7 | 29.5 | 36.7 |
| Sample c | 13.1 | 30.6 | 33.8 |

As we have already noted, if we compare sample a) and sample b) we find that, although they have approximately the same values of nonlinearity, sample a) SH emission is one order of magnitude higher than sample b) emission. The reason for this strong difference in the SH emission of sample a) and b) is clear if we resort to the Bloch vector analysis and in particular to the generalized momentum conservation condition (phase matching) for SH generation in periodic structures that in this case we write by resorting to the real part of the Bloch vector:

$$\pm \operatorname{Re} K_\beta(SH) \mp 2\operatorname{Re} K_\beta(FF) = mG_1 \qquad , \qquad (3)$$

where $G_1=2\pi/\Lambda$ is the first reciprocal lattice vector, $m$ is an integer that runs over all the positive and negative numbers including zero. The choice of the sign in front of the two Bloch vectors can be done independently each other giving therefore four cases:(+,-) forward SH/forward FF coupling, (-,+) backward SH /backward FF, (+,+) forward SH/backward FF, (-,-) backward SH/ forward FF. *From Eq.(3) and Table 1 we can realize that sample a) is phase matched with the first reciprocal lattice vector:* $\pm (\operatorname{Re} K_\beta(SH) + 2\operatorname{Re} K_\beta(FF)) \cong \pm G_1$, *while sample b) does not satisfy any of the conditions summarized in Eq.(3).* This finding by itself should confirm the robustness of the Bloch vector and clearly shows the signature of the periodicity even in finite, non-linear, dissipative systems. But there is more to say if we look at the power spectrum of the SH fields calculated inside the structure for the incident pump at 55 deg. In Fig.3 for $k_z>0$ the absolute maximum for all the figures corresponds to the spectral component of the forward SH emitted in the glass substrate ($n_s \sim 1.5$): $k_z = (2\pi/\lambda_{SH})\sqrt{n_s^2 - \sin^2(55^0)} \cong 19.7 \mu m$, while, obviously, for $k_z<0$ the absolute maximum corresponds to the backward SH emitted in air $k_z = -(2\pi/\lambda_{SH})\cos(55^0) \cong -13.5 \mu m$. A part from these two obvious components that give us information about the wave-vector of the SH field *outside* the sample, the other peaks are the most important because they tell us the spectral components of the field *inside* the sample. In Fig.3 we note that sample a) in the backward direction has two spectral peaks of emission centered respectively at $-K_\beta(SH)$ and $-G_1$ with the emission at $-K_\beta(SH)$ predominant. In the forward direction the peaks of emission are at $2K_\beta(FF)$ and $G_1$ with the peak of emission at $2K_\beta(FF)$ predominant. This peak at $2K_\beta(FF)$ is the signature of the bound (or phase-locked) SH [14,15], that, as well known [14], is generated at twice the wave-vector of the pump beam, in contrast with the standard "free SH" that is generated at the wave-vector of the SH. It is interesting to note that in this case the bound SH is generated at twice the pump Bloch vector, this is another strong evidence of the robustness of the Bloch vector that in this case is even able to rule the phase-locking mechanism. In sample b) and sample c) all the emissions fall under the same peak and so at this stage it is not possible to discriminate which one, if any, is favored. In order to asses which of the three possible emissions the periodicity is actually favoring, for sample b) and sample c) in particular, we have calculated the SH generated by three hypothetical structures (structure a), structure b) structure c)) with the same elementary cell and the same nonlinearities as our three samples, but a number of periods N=15. As we may expect, structure a) follows the same path traced by sample a), i.e. the forward SH is peaked at $2K_\beta(FF)$ and the backward SH is peaked at $-K_\beta(SH)$. Structure b) and structure c) in this case do discriminate between the possible emissions. It comes out that for sample b) in the backward direction the most favored emission is at $-G_1$.while in the forward direction the emission at $2K_\beta(FF)$ (bound SH) and at $G_1$ are approximately equally favored. Finally, for sample c) the forward/backward emissions at $\pm G_1$ are predominant.

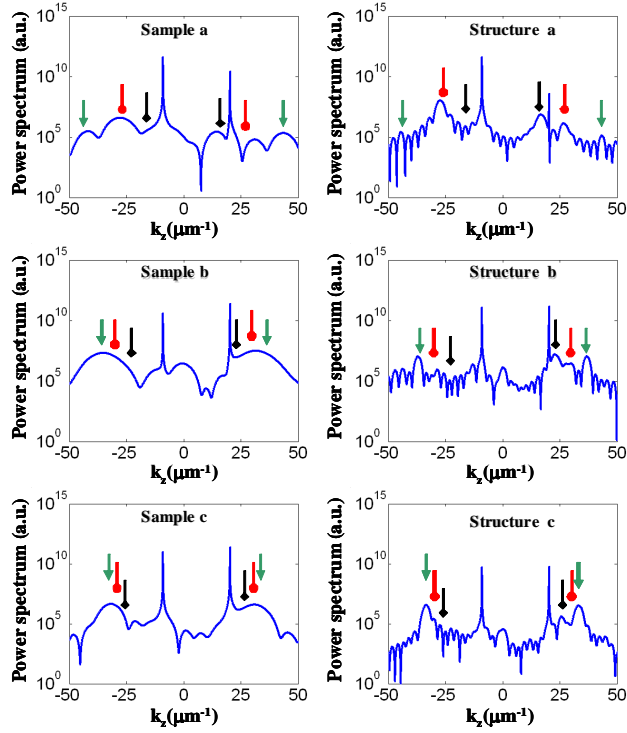

**Figure 3:** (Color online) Left-column: Power spectrum of the SH field for the three samples. Right column: Power spectrum of three hypothetical structures having the same elementary cell and same nonlinearities as the three samples, but different number of periods (N=15). Indicated the position of $\pm K_\beta(SH)$ (●), $\pm 2K_\beta(FF)$ (♦), $\pm G_1$ (▼).

*Conclusions*-In conclusion, we have given experimental evidences that the Bloch vector, as defined in Eq.(1), continues to play a key role even in nonlinear phenomena involving finite, dissipative systems, such as SH generation in MD structures. Although we have reduced our study to simple 1-D structures, we believe that similar considerations could be applied to multidimensional systems either. Nowadays that metal based periodic nanostructures are of central importance in the field of nano-photonics [16], our results clearly suggest that the Bloch vector still outstands among all the possible interpretative tools.

*Acknowledgments*-We thank M. Scalora, A. Heimbeck and M. Cappeddu for helpful discussions. G.D. and N.M thank the National Research Council for financial support.

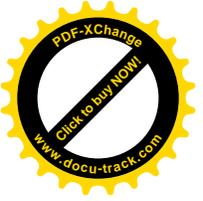
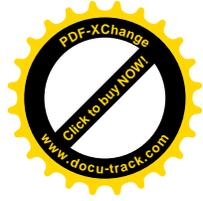

## References

[1] Ashcroft and Mermin, *"Solid State Physics"*, Holt, Rinehart and Winston, (1976); C. Kittel, *"Introduction to Solid State Physics"* Wiley, New York, (1996)

[2] A. Yariv, P. Yeh, *"Optical Waves in Crystals"* Wiley, New York, (1984); D.L. Mills, *"Nonlinear Optics"*, Springer, Berlin (1998); J.D. Joannopoulos, R.D. Meade, J.N. Winn, *"Photonic Crystals"*, Princeton University Press (1995)

[3] H.A. Macleod, *"Thin film optical filters"*, Institute of Physics Publishing (2001)

[4] P. Yeh, *"Optical Waves in Layered Media"*, Wiley, New York (1988)

[5] V. Kuzmiak and A.A. Maradudin, Phys. Rev. B **55**, 7427 (1997) and references therein

[6] M. Bergmair, M. Huber, and K. Hingerl, Appl. Phys. Lett. **89**, 081907 (2006)

[7] "Ag" in "Handbook of Optical constants of solids II" pp.737, E.D. Palik ed., Academic Press Inc., New York (1991).

[8] M.C. Larciprete et al., Phys Rev. A **77,** 013809 (2008)

[9] N. Bloembergen, R.K. Chang, S.S. Jha and C.H. Lee, Phys. Rev. **174**, 813 (1968).

[10] Y.R. Shen, " The Principles of Nonlinear Optics" Wiley New York (1984)

[11] N. Mattiucci et al., Phys. Rev. E **72**, 066612 (2005).

[12] D. Krause, C.W. Teplin, C.T. Rogers, J. Appl. Phys. **96,** 3626 (2004)

[13] J.L. Coutaz et al., J. Appl. Phys. **62**, 1529 (1986)

[14] J. Jerphagnon and S. K. Kurtz, J. Appl. Phys. **41**, 1667 (1970).

[15] M. Centini et al., Phys. Rev. Lett. **101**, 113905 (2008).

[16] Paras N. Prasad, *"Nanophotonics"*, Wiley New York (2004)